\documentclass[aps,prd,floats,floatfix,showpacs,twocolumn,twoside,preprintnumbers,superscriptaddress,nofootinbib]{revtex4}
\usepackage{graphicx,pstricks}
\usepackage{amsmath,amssymb}
\usepackage{amsfonts}

\newcommand{\ve}[1]{\mathbf{#1}}

\newcommand{\lcbar}{\hspace{1mm}\vspace{2mm}\bar{}\vspace{-2mm} \hspace{-1mm}\lambda_{C}}

\begin{document}

\title{Properties of the electron-positron
plasma  created  from   vacuum  in  a  strong  laser  field.
Quasiparticle excitations}
\author{D.~B.~Blaschke}
\email{blaschke@ift.uni.wroc.pl}
\affiliation{Institute for Theoretical Physics, University of Wroc{\l}aw,
50-204 Wroc{\l}aw, Poland}
\affiliation{Bogoliubov Laboratory for Theoretical Physics, Joint Institute
for Nuclear Research, RU - 141980 Dubna, Russia}
\affiliation{Fakult\"at f\"ur Physik, Universit\"at Bielefeld,
D-33615 Bielefeld, Germany}
\author{B.~K\"ampfer}
\affiliation{Helmholtz-Zentrum Dresden-Rossendorf, PF 510119, D-01314 Dresden, 
Germany}
\affiliation{Institut f\"ur Theoretische Physik, TU Dresden, D-01062 Dresden,
Germany}
\author{S.~M.~Schmidt}
\affiliation{Institute for Advanced Simulation, Forschungszentrum J\"ulich and 
JARA, D-52425 J\"ulich, Germany}
\affiliation{III. Physikalisches Institut B, Physikzentrum, RWTH Aachen,
D-52056 Aachen, Germany}
\author{A.~D.~Panferov}
\affiliation{Department of Mathematics, Saratov State University, RU-410026
Saratov, Russia}
\author{A.~V.~Prozorkevich}
\affiliation{Department of Physics, Saratov State University, RU-410026
Saratov, Russia}
\author{S.~A.~Smolyansky}
\email{smol@sgu.ru}
\affiliation{Department of Physics, Saratov State University, RU-410026
Saratov, Russia}
\date{\today}

\begin{abstract}
Solutions of a kinetic equation are investigated which describe, on a 
nonperturbative basis, the vacuum creation of quasiparticle electron-positron 
pairs due to a strong laser field.
The dependence of the quasiparticle electron (positron) distribution function  
and the particle number density is explored in a wide range of the laser 
radiation parameters, i.e., the wavelength $\lambda$ and amplitude of electric 
field strength $E_0$. 
Three domains are found: 
the domain of vacuum polarization effects where the  density of the $e^-e^+$ 
pairs is small (the ``calm valley''), and two accumulation domains in which 
the production rate of the $e^-e^+$ pairs is strongly increased and the 
$e^-e^+$ pair density can reach a significant value 
(the short wave length domain and the strong field one). 
In particular, the obtained results point to a complicated  short-distance
electromagnetic structure of the physical vacuum in the domain of
short wavelengths $\lambda \lesssim \lambda_{\rm acc} = \pi/m$.
For moderately strong fields $E_0 \lesssim E_c=m^2/e$, the accumulation regime 
can be realized where a plasma with a high density of $e^-e^+$ quasiparticles 
can be achieved.
In this domain of the field strengths and in the whole investigated range of 
wavelengths, an observation of the dynamical Schwinger effect can be 
facilitated.
\end{abstract}



\maketitle

\section{Introduction}

In the present work we investigate the response of the physical vacuum (PV)
of QED to the influence of a time dependent, strong periodic electric field
("laser field") in a wide range of field parameters, i.e., the field strength 
amplitude $E_0$ and the wavelength $\lambda$ (we use the natural units 
$\hbar = c = k_B = 1$, $k_B$ being the Boltzmann constant),
\begin{align}
  &0\leq E_0 \lesssim E_c = m^2/e, \label{01}\\
 & \pi/m = \pi\lcbar = \lambda_{\rm acc} \lesssim \lambda \leq \lambda_0 , 
 \label{02}
\end{align}
where $E_c$ is the so-called critical or Sauter-Schwinger field strength, and
$\lambda_C=2\pi \lcbar$ is the Compton wavelength of elementary 
leptons with mass $m$ and electric charge $e=|e|$ (we focus here on electrons 
and positrons); 
$\lambda_0 \sim 1~\mu$m \footnote{The actual value of this limit is determined
by the restrictions due to limited computing capabilities}.
To characterize the PV response we use either the distribution function or the
number density of quasiparticle electron-positron pairs (EPPs) created from 
the PV. 
Below we select three characteristic domains in the plane spanned by $E_0$ and
$\lambda$: 
the region of non-increasing creation (dubbed "calm valley") and two boundary
regions of accumulation, where the EPP density increases up to saturation 
(for sufficient duration of the field). 
The accumulation regions correspond to limiting high fields $E_0 \sim E_c$ 
(and $\lambda \gg \lambda_C$) or limiting short wavelengths 
$\lambda \sim \lambda_C$ (and $E_0 < E_c$).
Of course, there is  also a region where these accumulation domains do overlap.
Different mechanisms of vacuum creation act in these domains and provide the
corresponding features of the PV response.
The vacuum excitations in the calm valley can be described perturbatively
while nonperturbative approaches are necessary in both accumulation domains.
In the following, we consider the action of a periodical field. 
But the obtained results allow also to estimate the role of separate Fourier 
components of an external field with complicated spectral structure (as, e.g., 
in the cases of vacuum EPP creation by scattering of charged particles at high
energies \cite{Baur:2007df} and the dynamical Casimir effect 
\cite{Dodonov:2010zza}).

The methodical basis of our analysis is a system of kinetic equations  (KEs)
which is a nonperturbative consequence of QED in the presence of a spatially
homogeneous, time dependent electric field 
\cite{Kluger:1992,Schmidt:1998vi,Pervushin:2006vh}.
These KEs are intended for the description of the dynamical Schwinger
effect of vacuum particle creation in terms of quasiparticle vacuum
excitations, having the quantum numbers of the electron and positron in the 
presence of an external field (mass, charge, quasi-momentum and quasi-energy), 
which are called below as quasiparticles. 
After the ceasing of the laser pulse some of these quasiparticles become real 
electrons and positrons with their momentum and energy lying on the mass shell.

Thus, in the present work, vacuum polarization effects and vacuum particle 
creation in the time-dependent electric field of a standing wave formed
in the focal spot of counter-propagating strong laser fields are analyzed 
in the framework of the quasiparticle representation. 
Such a quasiparticle picture corresponds to the level of description with real 
EPPs in the out-channel. 
On the other hand, the information about quasiparticle distributions is useful 
for different estimates of the secondary observable effects such as the
generation of pair annihilation photons \cite{Blaschke:2005hs,Blaschke:2009uy,Blaschke:2010vs,Smolyansky:2010as, Blaschke:2011af, Blaschke:2011is}, 
the birefringence effect (e.g., \cite{DiPiazza:2006pr}) and so on. 
The additional investigation of the residual EPP surviving after the ceasing 
of the laser pulse will be given elsewhere.

The present work is organized as follows. 
In Sect.~\ref{sec:basic} the basic KE and the set of limitations for their 
applicability are presented.
The results of numerical investigations of the vacuum EPP creation kinetics
for the case of a linearly polarized laser field are discussed in 
Sect.~\ref{sec:PV} for different domains of the external field parameters in 
the ranges (\ref{01}) and (\ref{02}).
In the calm valley of the laser radiation parameters, the same results can be
reproduced analytically in the low density approximation by using the peculiar
perturbation theory (see Sect.~\ref{sec:low}).
The summary of the results is given and discussed in Sect.~\ref{sec:sum}.
Here the actuality of development of a nonperturbative kinetic theory of
vacuum EPP creation for the case of non-quasiclassical external 
electromagnetic fields is emphasized for cases where the quantum field
fluctuations are essential.
It is also shown that both accumulation domains are promising for an
experimental observation of the dynamical Schwinger effect.

\section{Basic kinetic equations} 
\label{sec:basic}

It is well known that vacuum particle creation in strong electromagnetic fields
is possible if one or both of the field invariants $I_{1} = E^2 - H^2$ and 
$I_{2} = {\vec{E}}{\vec{H}}$ are non-vanishing \cite{Sauter:1931zz,Heisenberg:1935qt,Schwinger:1951nm,Greiner:1985ce,Grib:1994}.
Such conditions can be realized, e.g., in the focal spot
of two or more counter-propagating laser beams, where the electric field
is spatially homogeneous over distances $\sim\lambda$ and time dependent.
For very strong fields $E_{0} \sim E_{c}$ one can then expect the creation of a
real electron-positron plasma
\cite{Grib:1994,Brezin:1970xf,Bulanov:2004de,Narozhny:2006,Fedotov:2006}
out of the electromagnetically polarized PV.
In the subcritical field regime $E_{0} \ll E_{c}$ the question is mainly about 
the creation of a short-lived quasiparticle EPP that exists during the 
course of the field action (the dynamical Schwinger effect) 
\cite{Blaschke:2008wf,Blaschke:2005hs}.
The possibilities for observing this kind of PV response with the help of 
secondary effects such as the radiation of annihilation photons are 
investigated intensively at the present time.

In this work, we study the PV response to a periodic model laser pulse with 
linear polarization $A^\mu (t) = (0 , 0 , 0 , A (t))$ (Hamiltonian gauge), 
where
\begin{equation}
A(t) = (E_{0}/\nu)\cos(\nu t), \quad E(t) = E_{0}\sin(\nu t)~.
\label{field}
\end{equation}
We are going to investigate the PV response as a function of the angular
frequency $\nu=2\pi/\lambda$ (or the wavelength $\lambda$) and the amplitude
$E_{0}$.
For this aim we use the exact nonperturbative KE of the non-Markovian type for 
the one-body EPP phase-space distribution function $f(\mathbf{p},t)$ obtained 
in \cite{Schmidt:1998vi},
\begin{equation}
\label{ke}
\dot f(\mathbf{p} ,t) = \Lambda(\mathbf{p}
,t)\int^t_0dt^{\prime} \Lambda(\mathbf{p} ,t^{\prime})
[\frac{1}{2}-f(\mathbf{p},t^{\prime})]\cos2\theta(t,t^{\prime}),
\end{equation}
where
\begin{equation}
 \Lambda(\mathbf{p},t) = e E(t)\varepsilon_{\bot}/\omega^{2}(\mathbf{p},t)
\label{3}\end{equation}
is the amplitude of the EPP excitations with the quasi-energy
$\omega(\mathbf{p} ,t) = \sqrt{\varepsilon^2_{\bot}(\mathbf{p}) +
(p_{\parallel}-eA(t))^2}$ and $\varepsilon_\bot = (m^2 + p^2_\bot)^{1/2}$ 
as the transverse energy; the quantity
\begin{equation}
\theta(t,t^\prime) = \int^t_{t^\prime} d\tau \ \omega(\mathbf{p} ,\tau)
\label{4}
\end{equation}
is the high frequency phase.
The distribution function in the quasiparticle representation is defined
relative to the in-vacuum state,
$f(\mathbf{p},t) = \langle in\mid a^{+}(\mathbf{p},t) a(\mathbf{p},t)\mid
in\rangle$, 
where  $a(\mathbf{p},t)$  and  $a^{+}(\mathbf{p},t)$  are  the  annihilation  
and  creation operators in the quasiparticle representation. 
The existence of the quasi-energy $\omega(\mathbf{p},t)$ suggests that 
the quasiparticle excitations are not on the mass shell.
For the generalization of the KE (\ref{ke}) to arbitrary electric field
polarization, see Refs.~\cite{Pervushin:2006vh,Filatov:2006,Filatov:2009xd}.

The KE (\ref{ke}) is equivalent to the system of ordinary differential
equations
\begin{eqnarray}
\dot f &=& \frac{1}{2}\Lambda u,\nonumber\\ 
\dot u &=& \Lambda (1 - 2 f) - 2 \omega v ,\\
\dot v &=& 2 \omega u \nonumber,
\label{5}
\end{eqnarray}
which is convenient for the subsequent numerical investigations. 
The functions $u(\mathbf{p},t)$, $v(\mathbf{p},t)$ describe the vacuum 
polarization effects.
Initial conditions at $t=0$ are $f=u=v=0$.

It is assumed that the laser electric field $E(t)$ (\ref{field}) is 
quasiclassical.
This means that the photon number with the frequency $\nu = 2 \pi / \lambda$ 
must be rather large in a volume of the order $\lambda^{3}$.
This condition is fulfilled \cite{Berestetskii,Ritus:1979} for
\begin{equation}
\label{qc}
E_0 \gg (\nu/2\pi)^2~,
\end{equation}
i.e., in the quasiclassical (QC) domain.
In the quantum (Q) domain for $E_0 \lesssim (\nu/2\pi)^2$ it is necessary to 
take into account the quantum fluctuations of the external  electromagnetic 
field and a corresponding generalization of the KE (\ref{ke}) is required. 

Below it will be shown that the features in the behavior of the PV response 
for the weak field case $E_0 \ll E_c$ become  apparent just in the Q domain, 
where the applicability of KE (\ref{ke}) breaks down. 
An exception is the strong field domain $E_0 \lesssim E_c$, where the 
inequality (\ref{qc}) is valid. 
However, in spite of this, we will investigate its solutions here assuming 
that the external field can be treated as some quasiclassical background. 
Thus, these extrapolated solutions have the character of a preliminary 
forecast.

\begin{figure}[!ht]
\includegraphics[width=0.48\textwidth]{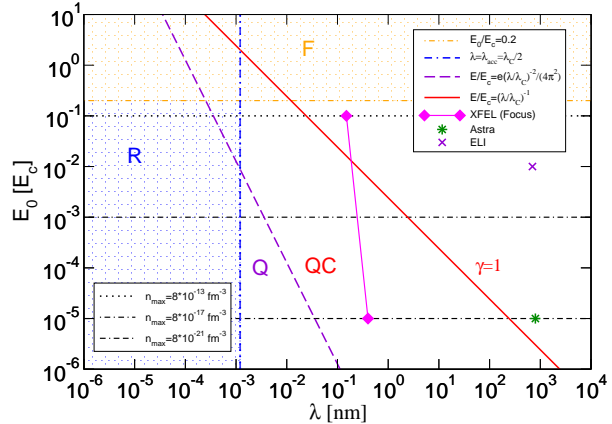}
\caption{(Color online) 
Electron-positron pair production in the ``landscape'' of field 
strength $E_0$ and wavelength $\lambda$ of laser collider experiments, 
with a few examples for comparison 
(XFEL \cite{Ringwald:2001ib}, Astra \cite{Gregori:2010uf}, ELI 
\cite{Korzhimanov:2011}).
The hatched areas show the regions of accumulation due to the strong field (F) 
or resonance (R) mechanisms, resp.
The black lines (dotted, dash-dotted, dash-double dotted) show three examples 
for peak pair densities $n_{\rm max}$.
The violet dashed line depicts the limit between the quantum (Q) and the 
quasiclassical (QC) domain which itself is limited by the red solid line 
(adiabaticity index $\gamma=1$).  
\label{fig:landscape}}
\end{figure}

The adiabaticity parameter \cite{Popov:2001,Popov:2004,Delone:1998}
\begin{equation}
\gamma = \frac{E_{c}}{E_{0}} \frac{\lambda_{c}}{\lambda}
\label{gamma}
\end{equation}
is introduced for separating the domains of influence of two
mechanisms of vacuum  particle creation: tunneling for $\gamma \ll 1$
and multiphoton for $\gamma \gg 1$ processes.
The point $\gamma = 1$ corresponds to the boundary curve  
$E_{c} \lambda_{c} = E_{0} \lambda$ 
in Fig.~\ref{fig:landscape} in the vicinity of which, however, no clear 
separation of these domains is possible.

It is assumed that the field (\ref{field}) is switched on at $t_{0} = 0$
(we ignore here the error brought in by such an instantaneous switching-on
\cite{Grib:1994,Blaschke:2010sgu}). 
This way of switching on the external periodical field is rather standard in 
the framework of the discussed problem \cite{Grib:1994,Brezin:1970xf}. 
Below we will consider also the method of adiabatic field switching on. 
Generally speaking, the obtained results depend on the way of the field 
switching on. 
This is due to the occurrence of high harmonics in the solution spectrum of 
the KE (\ref{ke}). 
As a rule, the high harmonics, which accompany the fast (non-adiabatic) 
switching on of a laser field, have no essential influence on the solution 
spectrum.

For a comparative investigation of EPP production as a function of the field  
parameters $\lambda$ and $E_{0}$ it is convenient to use some indicator 
characteristics. 
For this purpose we introduce here the maximal EPP number density 
$n_{\rm max}=n(t=T/4)$ which is related to the maximal distribution function  
$f_{\rm max}(\mathbf{p})=f(\mathbf{p},t=T/4)$ obtained from solving the KE
(\ref{ke}) by integration over the momentum space
\begin{equation}
n(t) = g\int\frac{d^{3}p}{(2\pi)^{3}}f(\mathbf{p},t)~,
\label{9}
\end{equation}
where $g = 4$ is the degeneracy factor, taking into 
account the spin and charge degrees of freedom.
This definition is used below in the case of absence of accumulation  effects  
for which the  number density EPP grows rapidly.
In Fig.~\ref{fig:landscape} the above discussion is summarized in the 
``landscape'' of the laser parameters $E_0$ and $\lambda$. 

\section{Short-distance electromagnetic structure of the PV} 
\label{sec:PV}

A numerical investigation of the KE (\ref{ke}) demonstrates the
complicated behavior of the distribution  function which shows a
series of qualitative modifications where the laser
radiation parameters $E_0$ and $\lambda$ are varied in the boundaries 
(\ref{01}) and (\ref{02}).
It is important to note that the character of the distribution function 
evolution depends strongly on the selection of the momentum representation:
$f(\textbf{p},t)$ accomplishes some oscillations swinging along the 
$p_{\parallel}$ axis 
\footnote{This follows from the definition of the kinematic momentum 
$P=p_{\shortparallel}-eA(t)$ and the construction (\ref{field}) of the laser 
field.} (the direction of the field (\ref{field})) as a whole
with the amplitude $1/\gamma$ 
simultaneously with the "breathing" mode, when its amplitude and form are 
altered. 
The transition to the kinematic momentum eliminates these oscillations and
keeps the "breathing" mode only.

\begin{figure}[!ht]
\includegraphics[width=0.48\textwidth]{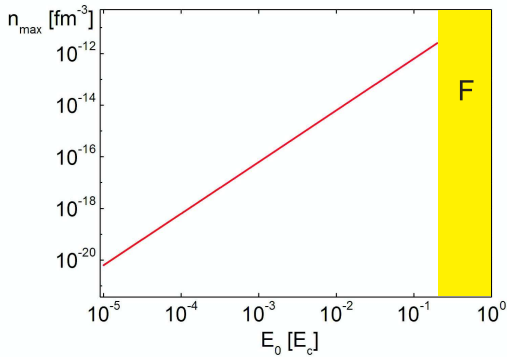}
\caption{The dependence of $n_{\rm max}$ on the field strength $E_0$ in
the calm valley.
\label{fig:nmax-E0}}
\end{figure}

The characteristic domains of dynamical behavior of the distribution function 
$f(\textbf{p},t)$ and the number density of EPPs (\ref{9})  in the full range 
of the laser radiation parameters (\ref{01}) and (\ref{02}) are shown in 
Fig.~\ref{fig:landscape} to be discussed step by step in what follows. 
The region outside the hatched accumulation domains due to strong field (F) 
and resonance (R) mechanisms of pair production is the calm valley. 
The boundary between quasiclassical (QC) and quantum (Q) domains of the 
electric field (\ref{field}) is given by $E_0=1/\lambda^2$ and the line  
$\gamma  =  1$ separates the multiphoton and tunneling domains.
Surprisingly, $n_{\rm max}$ is $\lambda$ independent 
practically in the whole calm valley, i.e. outside the accumulation domains 
R and F. Its dependence on the strength of electric field $E_0$ is shown in
Fig.~\ref{fig:nmax-E0}. 
The symbols on Fig.~\ref{fig:landscape} depict also the basic parameters  
in the focal spot of the existing \cite{Liesfeld:2005,Gregori:2010uf}
and planned \cite{Ringwald:2001ib,DiPiazza:2011tq,Korzhimanov:2011} laser 
systems as, e.g., Astra, XFEL and ELI.

In the following subsections we discuss the three characteristic domains of 
Fig.~\ref{fig:landscape} separately.

\subsection{The calm valley domain} 

This domain corresponds to the most simple behaviour of the distribution 
function characterized by the absence of an appreciable production rate for
EPP when averaged over the period of field oscillation.

\begin{figure}[!ht]
\includegraphics[width=0.48\textwidth]{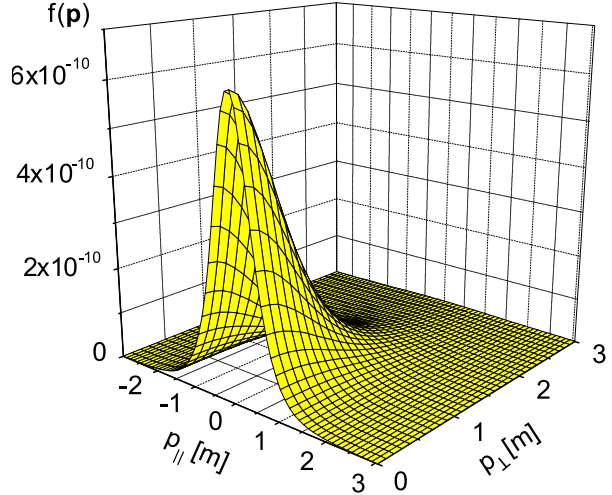}
\caption{The EPP distribution function $f(p_{\bot},p_{\parallel})$
for the laser radiation with  $\lambda = 1$ nm and $E = 10^{-4}E_c$
at the moment $t=T/4$. \label{fig_hill}}
\end{figure}

It is interesting to consider the initial behaviour of the distribution 
function $f(\textbf{p})$ immediately after the field is switched on.
In Fig.~\ref{fig_hill} we present the typical form of $f(\textbf{p})$ 
at the time $t=T/4$, where $T=2\pi/\nu$ is the period of the laser field 
oscillation. 

\begin{figure}[!ht]
\includegraphics[width=0.48\textwidth]{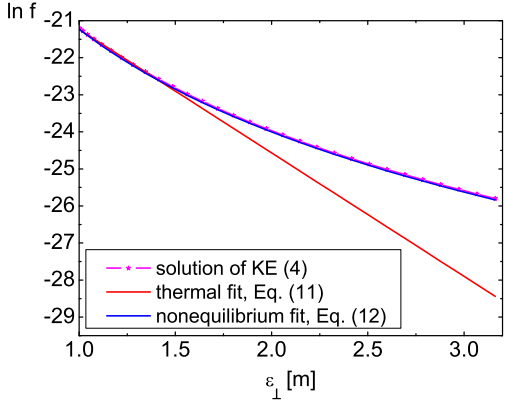}\\
\includegraphics[width=0.48\textwidth]{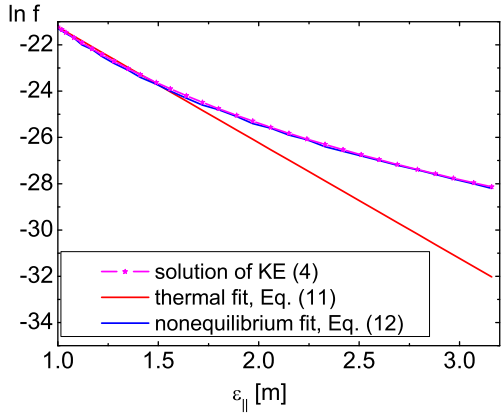}
\caption{Fitting the KE solution $f(\textbf{p},t=T/4)$ of Fig.~\ref{fig_hill}
(dash-dotted line) with an anisotropic equilibrium Boltzmann distribution 
(\ref{feq}) with $T_\perp =0.3~m$ (upper panel) 
and $T_\parallel =0.19~m$ (lower panel) as well as with a power-like 
distribution (\ref{om6}).
\label{fig_feq}}
\end{figure}

This distribution is anisotropic and can be approximated by a two-temperature 
Boltzmann distribution
 \begin{equation}
\label{feq}
f_{\rm eq}(p_\perp,p_\parallel) \sim 
\exp{\{-\varepsilon_\perp /T_\perp -\varepsilon_\parallel /T_\parallel \}}~,
\end{equation}
where $T_\perp$ is the transversal and $T_\parallel$ the longitudinal 
temperature while $\varepsilon_\parallel(\textbf{p})=\sqrt{m^2+p_\parallel^2}$ 
is the longitudinal energy.
Fig.~\ref{fig_feq} demonstrates the similarity of the distributions 
$f(\mathbf{p})$  and $f_{eq}(\mathbf{p})$  (\ref{feq}) and the relation for 
the temperatures $\xi =  T_\parallel/T_\perp$. 
However, a more satisfactory approximation for the pair distribution function
is
\begin{equation}
\label{om6}
f_{\rm in}(\textbf{p})\sim\varepsilon^2_\perp(p_\perp)/\omega^6_0(\textbf{p})~.
\end{equation}
The difference between the distributions (\ref{feq}) and (\ref{om6}) appears 
in the high-energy tails.
The estimate (\ref{om6}) can be obtained on the basis of the KE (\ref{ke}) 
in the low density approximation in the multiphoton domain $\gamma \gg 1$ 
(see also Sect.~\ref{sec:low}). 
In the framework of the approximation (\ref{feq}) the temperature
parameters $T_\perp$ and $T_\parallel$ are universal and do not depend on the 
laser field parameters $\lambda$ and $E_0$. 
In these assumptions the distributions (\ref{feq}) and (\ref{om6}) are 
independent of the wavelength $\lambda$ and proportional to $E_0^2$. 
Moreover, in the initial stage these estimates remain valid regardless of the
form of the electric field pulse.
This holds in particular for the Sauter potential, where the solution is well 
known \cite{Nikishov:1969tt,Narozhnyi:1970uv,Grib:1994}.

The excitation mechanism of the anisotropic spectrum of EPPs in the early 
stage of the evolution is fully determined by the dynamics of vacuum creation 
and, specifically, by the structure of the amplitude (\ref{3}). 
There the anisotropy effects are represented by both, the transverse energy
$\varepsilon_\perp(p_\perp)$ and the quasienergy $\omega(\mathbf{p},t)$. 
The same anisotropy factor $\xi = 2/3$ can be obtained on the basis of the 
distribution (\ref{om6}), when the relation of derivatives of the projections 
$f_{in}(p_\parallel, p_\perp=0)$ and  $f_{in}(p_\parallel =0, p_\perp)$ with 
respect to $\varepsilon_\parallel$ and $\varepsilon_\perp$ is considered.

Thus, in the case of a short pulse, the anisotropic distributions (\ref{feq}) 
and (\ref{om6}) will lead to an elliptic flow of EPPs, which is compressed in 
the direction of the electric field.

\begin{figure}[!h]\centering
\includegraphics[width=0.48\textwidth]{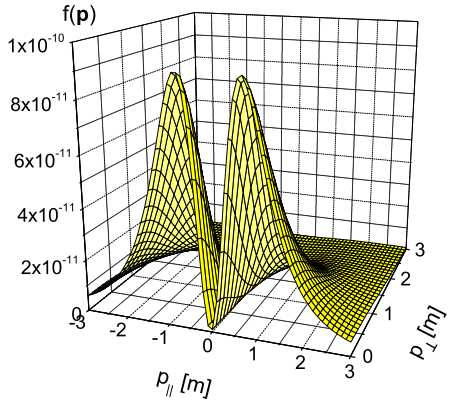}
\caption{Momentum space distribution of bosons for the same conditions as in  
Fig.~\ref{fig_hill}.
\label{fig_bose}}
\end{figure}

The estimates (\ref{feq}) and (\ref{om6}) are modified when other mechanisms 
of vacuum creation of fermions are realized (see, e.g., \cite{Filatov:2007ha}).
This can lead to a change of the anisotropy factor $\xi$. 
Considering, instead of fermions, the case of bosons, the situation changes
radically: 
the equilibrium-like distributions of the type (\ref{feq}) and  (\ref{om6}) 
are replaced by the ones for strong nonequilibrium \cite{Schmidt:1998zh}, 
see Fig.~\ref{fig_bose}.

Thus, one can say that the initial distribution of the particles created from 
vacuum is well defined in the general case. 
In particular, the conclusion about the existence of equilibrium-like 
universal distributions  can become important for the theory of quark-gluon 
plasma generation in heavy-ion collisions.

A few remarks are in order. 
Apparently, the temperature parameters $T_\perp$ and $T_\parallel$ can not be 
interpreted as the Unruh temperature $T_U=a/2\pi$ \cite{Castorina:2007eb}, 
since the acceleration $a=-(e/m) \dot{A}(t) \sim E_0$ would imply a field 
dependence of $T_\perp$ and $T_\parallel$ which is not observed in our 
numerical investigations. 
Some anisotropic initial distributions of particles created from 
vacuum under the action of an electric field have been introduced long ago 
(see, e.g., \cite{Ruffini:2009hg}), but they have artificial character.

While a parametrization of the distribution function resulting from the 
KE (\ref{ke}) by a Boltzmann distribution may be a covenient characterization
at certain time instances and in restricted intervals of momenta or energies,
the concept of temperature in the usual thermodynamical or statistical sense
is obviously not applicable. In the present example, it is the strong 
anisotropy which signals a striking off-equilibrium situation not accessible by
equilibrium or near-equilibrium thermodynamics.

\begin{figure}[!ht]
\includegraphics[width=0.48\textwidth]{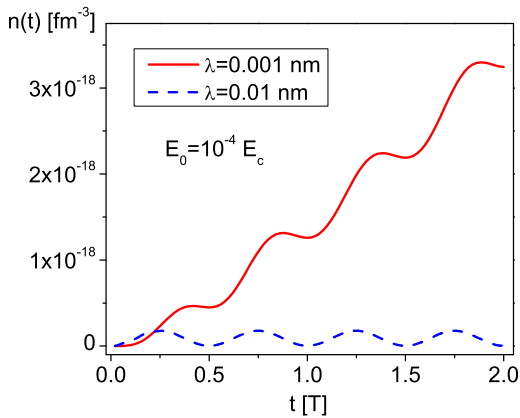}
\caption{Time dependence of density $n(t)$ in the weak field regime 
evidencing the transition from the oscillating mode for $\lambda =0.01$ nm 
to the linearly growing (in the mean) mode at $\lambda =0.001$ nm.
The latter behaviour is termed here as ``accumulation''.
\label{fig_accum}}
\end{figure}

Such a simplifying picture of the phenomena in the calm valley is valid 
approximately in the initial stage of the process for $t \lesssim T$. 
The high frequency harmonics become essential at $t \gg T$  and the structure 
of the distribution function gets very complex. 
However, the amplitude estimate remains valid and the EPP production rate 
averaged over a period (i.e., the pair creation per unit time) is absent, 
$<\dot{f}>_T\approx 0$.
The dependence $n_{\rm max}(E_0)$ as depicted in Fig.~\ref{fig:nmax-E0}
reveals that in the whole calm valley domain holds $n_{\rm max}(E_0)\sim E^2_0$
up to the strong field accumulation region (the shaded area ``F'' in that 
figure) where the numerical calculation is complicated.
The distribution function $f(\textbf{p},t)$ shows a breathing
oscillation with the amplitude $f_{\rm max}$ and a frequency being twice that
of the laser field (see Fig.~\ref{fig_accum}, dashed line). 

These features of the distribution function behavior are reproduced very well 
analytically on the quasiparticle level in Sect.~\ref{sec:low}. 
The residual number density of EPP is very small in comparison with the 
breathing mode.

The calm valley is bounded from above and to the left by the two accumulation
domains F and R (see Fig.~\ref{fig:landscape}), where the amplitude of  the 
distribution function increases with the lapse of time and the averaged EPP 
production rate $<\dot{f}>_T$ becomes appreciable.
Apparently, a glimpse of accumulation effect is presented in the calm valley
too. 
However, this effect is negligibly small for $\nu \ll m$ and $E_0 \ll E_c$. 
Thus, the accumulation effect becomes dominant in the F and R domains only.

\subsection{The resonance accumulation domain}
\label{ssec:res}

\begin{figure}[!ht]
\includegraphics[width=0.48\textwidth]{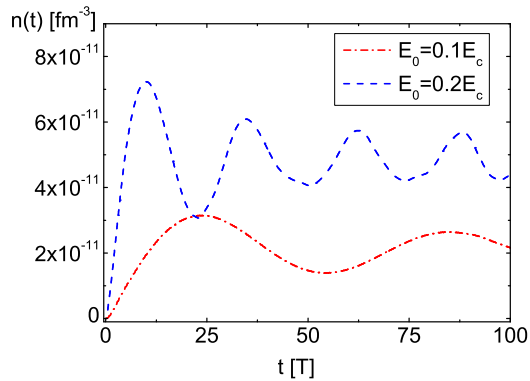}
\caption{The limitation of accumulation due to the saturation of
$f(\textbf{p})$ at  large times for $\lambda =0.001$ nm.
The saturation is achieved faster for stronger fields.
\label{degen}}
\end{figure}

This domain starts in the short wavelength domain from the point
$\lambda_{\rm acc} \approx \pi\lcbar=0.0012$ nm,
corresponding to the energy $\nu = 2m$, which peaks out from the 
monochromatic external field for creation of an EPP. 
We will denote this process as one-photon pair creation. 
Below we will use this term in the analysis of solutions of the KE (\ref{ke}), 
for which the spatial inhomogeneity effects are negligible.
Here, in the point $\lambda_{\rm acc}$ the accumulation mechanism is switched 
on sharply.
As an example, the initial growth of the density in the course of time is
depicted in Fig.~\ref{fig_accum} for $\lambda=0.001$ nm and $E_0= 10^{-4}~E_c$.
At the later stages the degeneration effect developes.
Here the distribution function reaches its maximal value and after that it
performs oscillations which are damped asymptotically (see Fig.~\ref{degen}).
In addition, the $2\nu$-dependence of the breathing mode is conserved in the
R domain. 

\begin{figure}[!ht]
\includegraphics[width=0.48\textwidth]{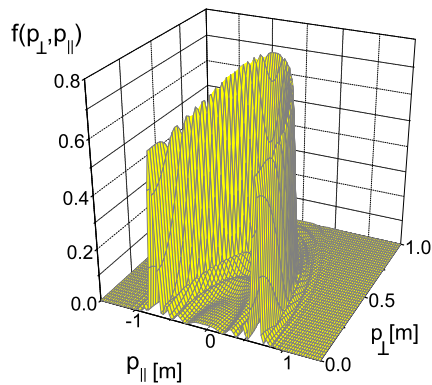}
\caption{The distribution function $f(p_{\bot},p_{\parallel})$ in the R domain 
at time $t\gg T$. 
Most of pairs are seen within a thin spherical layer centered at the origin of 
the coordinates.
\label{comp1}}
\end{figure}

\begin{figure}[!ht]
\includegraphics[width=0.48\textwidth]{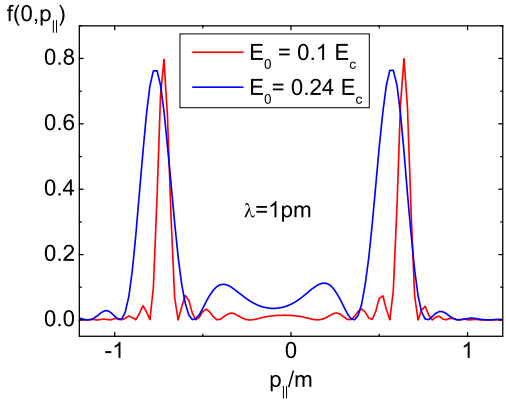}
\caption{The $f(0,p_{\parallel})$ dependence in the R domain
at large times after  saturation for different field strengths.
\label{comp2}}
\end{figure}

The shape of the distribution function is changes rapidly.
Later on, for $\lambda \leq \lambda_{\rm acc}$, it leads to a collapse of the 
distribution function in a thin spherical layer (EPP bubble, see 
Figs.~\ref{comp1} and \ref{comp2}).
The thickness $\Delta p$ and the size of this layer depend on the field 
parameters $\lambda$ and $E_0$.
For example, for $\lambda=0.001$ nm and $E_0=0.24~E_c$ we have
$\Delta p \sim 0.1~m$.
With growing field strength at a fixed frequency the maximum value 
$f_{\rm max}(\textbf{p})$ increases in the degeneration condition, 
$f_{\rm max}(\textbf{p}) \to 1$.
The occupation of the EPP bubble increases also (Fig.~\ref{comp2}) under this
condition (appearance of some thin wave structure arises within the bubble).
On the other hand, the decrease of the wavelength in the domain
$\lambda \lesssim \lambda_{acc}$ does not change the distribution picture
essentially (see Fig.~\ref{fig:f(p,lambda)}).
\begin{figure}[!ht]
\includegraphics[width=0.48\textwidth]{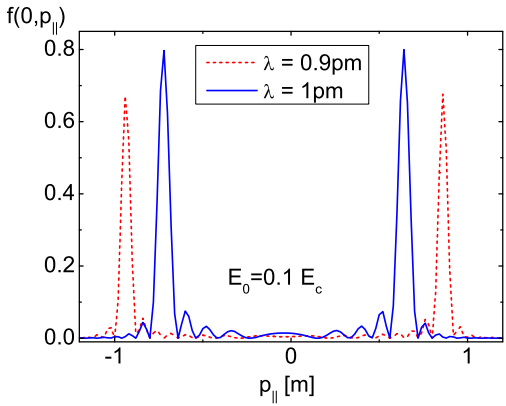}
\caption{The $f(0,p_{\parallel})$ dependence in the R domain
at large times after  saturation for different wavelengths.
\label{fig:f(p,lambda)}}
\end{figure}
The radius of the EPP bubble in momentum space is equal to
\begin{equation}
\label{radius}
p_{\rm bub}= \sqrt{(\pi/\lambda)^2 - m^2}~~
{\rm for}~~\lambda \leq \lambda_{\rm acc}~.
\end{equation}
For $\nu \to 2m$, this radius goes to zero, $p_{\rm bub}\to 0$ (it corresponds 
to the condition $\nu = 2 m$ or the point $\lambda=\lambda_{acc}$). 
Decreasing the wavelength is accomplished by the growth of the EPP bubble size 
(see Fig.~\ref{fig:f(p,lambda)}).
The maximal value of the number density of the vacuum bubble in the saturation 
state (see Fig.~\ref{fig:rd1}) and the time period necessary for achieving the 
saturation (see Fig.~\ref{fig:rd2}) show that the first passage time of the 
saturation grows indefinitely with decreasing $\lambda < \lambda_{acc}$.
\begin{figure}[!ht]
\includegraphics[width=0.48\textwidth]{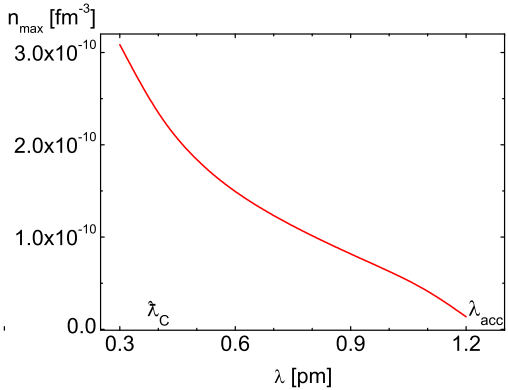}
\caption{The dependence of $n_{max}$ on wavelength $\lambda$ in the R domain 
for $E_0=0.1~E_c$. 
\label{fig:rd1}}
\end{figure}
\begin{figure}[!ht]
\includegraphics[width=0.48\textwidth]{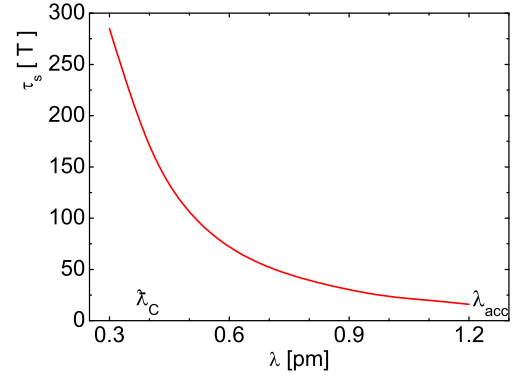}
\caption{The dependence of the saturation time $\tau_s$ on the wavelength 
$\lambda$ in the R domain for $E_0=0.1~E_c$ 
\label{fig:rd2}}
\end{figure}
From Fig.~\ref{fig:landscape} it follows that the point  $\lambda_{\rm acc}$ 
lies on the left of the critical line $\gamma=1$ in the domain $\gamma \gg 1$, 
where the multiphoton mechanism of EPP excitation is operative. 
Apparently, one can expect that the multi-photon process can be changed to a 
few-photon one in the lower left part of this figure.
As the result, the few-photon domain is limited by the two- and one-photon 
processes.
For weak fields $E \ll E_c$, the one-photon process of the EPP creation 
dominates (this conclusion is confirmed by the analytical calculations in 
Sect.~\ref{sec:low}). 
The two-photon pair creation process (the inverse of the Breit-Wheeler process 
\cite{Berestetskii,Ruffini:2009hg,854971}) is switched on for rather strong 
field $E_0 \sim E_c$ beginning with 
$\lambda_{2\gamma}=\lambda_C = 2\lambda_{\rm acc}$, see 
Fig.~\ref{fig:transient}.
\begin{figure}[!ht]
\includegraphics[width=0.48\textwidth]{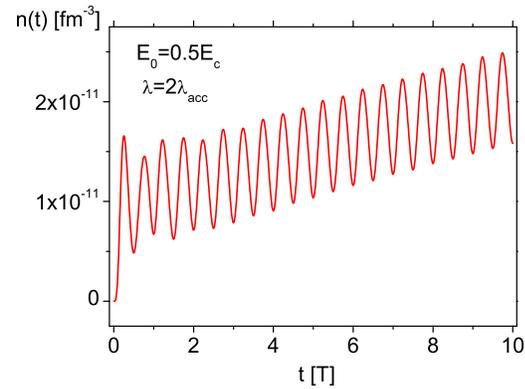}
\caption{Transient region from calm valley to F accumulation: the
characteristic for F domain with linear growth of mean density is here
caused by the subcritical field $E_0 = 0.5 E_c$, at the wavelength 
for the opening of the two-photon production channel 
$\lambda=\lambda_C=2\lambda_{\rm acc}$.
\label{fig:transient}}
\end{figure}
This domain is characterized by the fast growth of the EPP density and it is
contained in the F accumulation domain.
Surprisingly, the KE (\ref{ke}) is sensitive to this effect which is valid in
the case of a quantized electromagnetic field.
Let us remark that the one-photon mechanism of EPP excitation prolongs to act
in all domains of short wavelengths  $\lambda < \lambda_{\rm acc}$.


After having discussed implications of the purely periodic field (\ref{field})
we turn now to two other examples of the time dependence.
The initial condition is $f(t\to -\infty) = 0$.

For the first time, some features in the behaviour of the EPP 
distribution function in the domain of extremely short wavelengths 
($\lambda \sim \lambda_C$) were observed theoretically for the residual 
EPP at $t \to \infty$ in the framework of different approaches 
taking into account a spatial inhomogeneity of the external electric field
in the works \cite{Hebenstreit:2011wk,Gies:2005bz,Dunne:2006ur}. 
In the present work we investigate the behavior of the quasiparticle EPP 
within a period of the laser pulse action. 
It is well known that the properties of the real (residual) and 
quasiparticle EPP differ strongly (see, e.g., Fig.~14). 
It is possible that just this feature explains the difference between the 
results of the present work and Refs.~ 
\cite{Hebenstreit:2011wk,Gies:2005bz,Dunne:2006ur} regarding the short distance 
behaviour of the EPP. 
In the framework of the formalism based on the KE (\ref{ke}) such a comparison 
shall be performed within a separate work subsequent to this.
On the other hand, as pointed out above, the Q domain
(Fig.~\ref{fig:landscape}) is the domain of unreliable predictions obtained 
on the basis of an extrapolation of the domain of applicability of the KE 
(\ref{ke}) limited by the quasiclassical external electric fields where the 
electromagnetic fluctuations are inessential.

\subsubsection{Gaussian envelope}

\begin{figure}[!ht]
\includegraphics[width=0.48\textwidth]{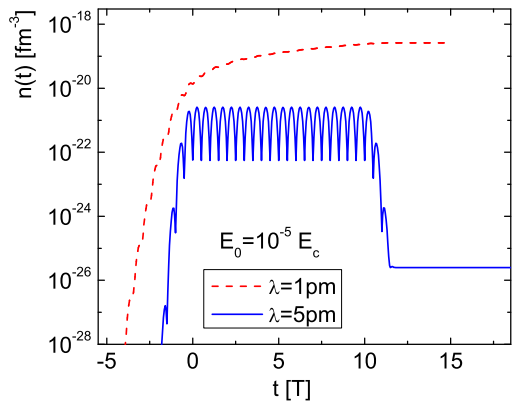}
\caption{Transition from the oscillating mode (for $\lambda = 5$ pm) to the 
accumulation mode (for $\lambda = 1$ pm) for the pulse (\ref{10a}) 
with $\tau_G = 10$.
\label{gauss}}
\end{figure}

Let us investigate first the PV response to the action of a pulsed field with 
Gaussian temporal envelope
\begin{equation}
\label{10a}
E(t) = E_0 \sin{ (\nu t)} \exp{ (-t^2/\tau_G^2)}~.
\end{equation}
The comparison  of the accumulation effects for the periodical field
(\ref{field}) with $\lambda = 0.001$ nm (dashed line) and the corresponding
field pulse (\ref{10a}) with $\tau_G =10$ T is shown in Fig.~\ref{gauss}.
It allows to draw the conclusion that the form of the field pulse is rather
essential but it does not change qualitatively the picture of the effect.
The switching-off  process of the pulse (\ref{10a}) is accompanied  by a 
stepwise reduction of the EPP density down to some residual level, see
Fig.~\ref{gauss}, which should be compared with Fig.~\ref{fig_accum}.
Let us remark that in the regime of saturation the residual EPP density can
surpass considerably the amplitude of the breathing oscillations while in the 
calm valley the situation is opposite.

\subsubsection{Sauter pulse}

Now we consider action of a smooth pulse field with the Sauter potential
\begin{eqnarray}
\label{10b}
A(t) &=& -E_0 \tau_S \tanh{(t/\tau_S)}~,\\ 
E(t) &=& E_0 /\cosh^2{(t/\tau_S)}~.
\label{10c}
\end{eqnarray}

\begin{figure}[!ht]
\includegraphics[width=0.48\textwidth]{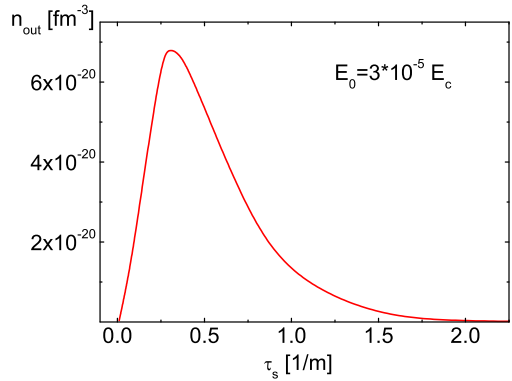}
\caption{The residual pair density $n_{\rm out}$ 
as a function of $\tau_{S}$ for the pulsed field (\ref{10c}). 
\label{fig_res1}}
\end{figure}

\begin{figure}[!ht]
\includegraphics[width=0.48\textwidth]{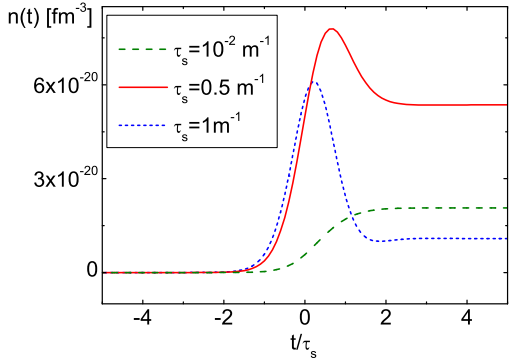}
\caption{The time dependence of pair number density for the pulsed field 
(\ref{10c}) for three values of the pulse length parameter $\tau_{S}$. 
\label{fig_res2}}
\end{figure}

Similar to the field (\ref{10a}) one can clearly define here the residual 
density $n_{\rm out}=n(t\to \infty)$.
The typical  resonance picture is revealed here  with the maximum  at 
$\tau_S \approx 1/2m$, see Figs.~\ref{fig_res1} and \ref{fig_res2}.
For $\tau_S \ll 1/2m$ a sharp decrease of the PV response takes place and the 
vacuum EPP creation practically ceases.
This result corresponds to the conclusion of Ref.~\cite{Hebenstreit:2011wk}.
The characteristic  value $\tau_S = 1/2m$ allows to speak here about the 
influence of one-photon pair production process on the level of the kinetic 
description. 
This feature in the solution  of the KE (\ref{ke}) with the  potential 
(\ref{10b}) can be confirmed also by an analysis of the exact solution of the 
problem \cite{Narozhnyi:1970uv}.

The accumulation effect in the R domain can be explained if one takes into
account the presence in the solution of the KE (\ref{ke}) of a parametric 
resonance in the neighborhood $\nu =2m$ and of all combinational frequencies 
$n\nu$ and $2m l$, where $n$ and $l$ are the integers and the factor 2 is 
stipulated by the structure of the high frequency phase (\ref{4}). 
This is compatible with a parity odd  distribution function $f(\mathbf{p},t)$ 
under time reversal. 
It leads to the appearance of contributions of the type
\begin{equation}
\label{sinsin}
    \frac{1}{\nu-2m}\  \sin{(\nu-2m)t} \, \cdot \, \sin{(\nu+2m)t}~.
\end{equation}
For $\nu \sim 2m$ a secular term appears here that results in a linear growth 
of the distribution function. 
The subsequent evolution is accompanied by a growth of the EPP number density 
and saturation as a result of the action of the statistical factor in the KE 
(\ref{ke}). 
Just this picture is observed for the numerical solution in the R-accumulation 
domain. 
The given interpretation was generated by the perturbation theory in the low 
density approximation, see Sect.~\ref{sec:low}.

Let us remark also that the motion along the $\lambda$-axis on the side of
short waves is accompanied by a sequential replacement of the vacuum creation
mechanisms: the tunneling mechanism acts in the limit of stationarity
($\lambda \to \infty$) and slowly alternating field and requires an infinite 
photon number from the external field reservoir. 
This mechanism is then replaced by the multiphoton ($\gamma \gg 1$) and 
few-photon ones and finally turns into the one- and two-photon pair production.
It is necessary to underline that the conception of the "photon" is considered 
here in the framework of the accepted model of the spatially homogeneous 
electric field acting in the focal spot of counter propagating laser fields
with the linear size $\sim \lambda$ that allows to use in the description of 
the absorption processes the energetic condition only (e.g., $\nu = 2m$ for 
the one-photon $e^-e^+$ production process).

\subsection{The strong field accumulation domain} 

\begin{figure}[!ht]
  \includegraphics[width=0.48\textwidth]{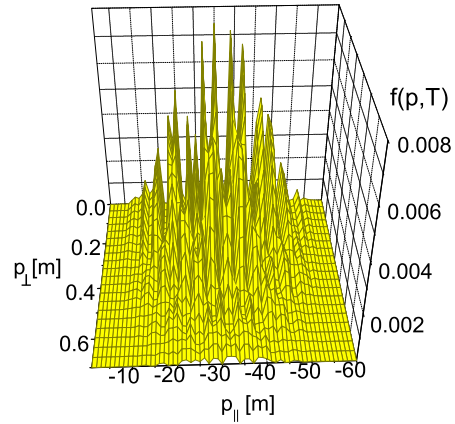}\\
  \includegraphics[width=0.48\textwidth]{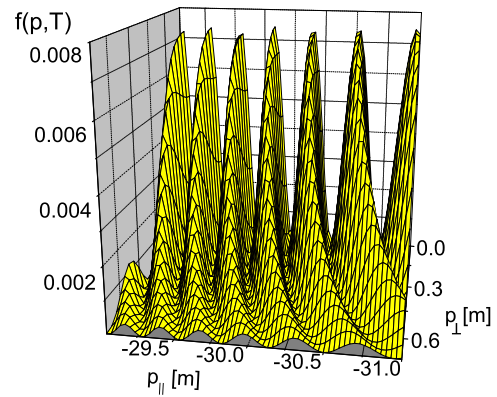}
  \caption{The distribution function in the F accumulation domain at $t=T$ 
($E_0 = 0.5~E_c,~\lambda = 0.15$ nm). 
Upper panel: general view; lower panel: detailed view highlighting the 
concentric dense layers. }
\label{fig:wafer}
\end{figure}

After the interlude on finite pulses let us return to the discussion of 
further effects related to the periodic field (\ref{field}).
For the first time, the accumulation effect in a strong laser field 
(domain F in Fig.~\ref{fig:landscape}) was discovered in
Ref.~\cite{Alkofer:2001ik} and further explored in \cite{Roberts:2002py} 
on the basis of a numerical solution of the KE (\ref{ke}).
The accumulation domain begins here with a rather high field strength 
$E_0 \approx 0.2 E_c$. 
The smooth breathing mode with a smooth phase space distribution
(Fig.~\ref{fig_hill}) is replaced here by a strongly fragmented one with 
pronounced structures (Fig.~\ref{fig:wafer}).
Apparently, in the long wavelength limit the usual accumulation mechanism acts 
here.
For the transition to the observable domain, the electron and positron must
gather within a spatial extent of one Compton wavelength, an energy which is 
comparable to the energy gap, i.e. $eE_c\lcbar=2m$.
In the multi- and few-photon domains the other mechanism is switched on where
the EPP creation is a result of the simultaneous absorption of $n$ identical
photons from the "photon reservoir"  of the laser field.
The probability for such kind of process, in principle, can be taken into 
account on the basis of an analogy with the theory of
multiphoton ionization of atoms given, e.g., in Ref.~\cite{Delone:1998}.

The intersection region of the two accumulation domains R and F (the left top 
corner on Fig.~\ref{fig:landscape}) is especially interesting. 
Here the nonlinear superposition of two accumulation effects can lead to some 
new features in the vacuum EPP production which have been investigated in 
recent years under the name of ``dynamically assisted Schwinger mechanism''
\cite{Schutzhold:2008pz,Orthaber:2011cm,Fey:2011if}. 
It is important that the applicability of the KE (\ref{ke}) is not violated in 
this domain.

Thus, the accumulation processes in the R and F domains are the processes of 
the EPP excitation up to the saturation state and next increasing of EPP 
density in this state.
Let us note that the vacuum response reveals the symmetry of the
states  in a weak ($E_0\ll E_c$) and a strong ($E_0 \sim E_c$)
field, respectively: under the substitution $f=1-\tilde{f}$ in the KE 
(\ref{ke}) it does not change its form and the case $f\ll 1$ is similar to the 
case $\tilde{f} \ll 1$.

\section{Low density approximation} 
\label{sec:low}

The numerical results discussed in Sect.~\ref{sec:PV} can be confirmed 
analytically in the multiphoton domain ($\gamma \gg 1$) of the calm valley.
From the KE (\ref{ke}) it follows in the low density approximation $f \ll 1$
corresponding to $E_{0} \ll E_{c}$ \cite{Blaschke:2010sgu} that
\begin{equation}
f(\mathbf{p} ,t) = \frac{1}{4} \bigg|\int^t_{t_{0}}dt^\prime
\lambda(\mathbf{p} ,t^\prime)
{\rm e}^{2i\theta(t^\prime ,t_0)}\bigg|^{2}.
\label{ld1}
\end{equation}
This formula solves correctly the initial value problem in the KE (\ref{ke}) 
providing a vanishing distribution function both at $t\to t_0$ and $E_0\to 0$. In the leading approximation 
$\omega(\mathbf{p},t) \rightarrow \omega_0(\mathbf{p}) = 
\sqrt{m^2 + p^2}$ , corresponding to $\gamma \gg 1 $, after the time 
integration, one arrives at
\begin{equation}
f(\mathbf{p},t) = f_{(0)}(\mathbf{p}) + \frac{1}{2} f_{(2)}(\mathbf{p})[1-\cos2\nu t],
\label{11}
\end{equation}
where ($\alpha =e^2/4\pi$)
\begin{eqnarray}
f_{(0)}(\mathbf{p}) &=&
\pi \alpha E^{2}_{0}
\frac{\nu^{2}\varepsilon^{2}_{\bot}}{\omega_0^{4}[4 \omega_0^{2}-\nu^{2}]^{2}},
\label{12}\\
f_{(2)}(\mathbf{p}) &=&
{\pi \alpha E^{2}_{0}} \frac{\varepsilon^{2}_{\bot}}
{\omega_0^{4} |4 \omega_0^{2} - \nu^{2}|}.
\label{13}
\end{eqnarray}
In these calculations, the assumption of adiabatically switching-on the 
electric fields (1) at $t_{0} \rightarrow -\infty$ was made.

The leading role of the second harmonics (\ref{11}) was detected in
the numerical solutions of the KE (\ref{ke}) (see Sect.~\ref{sec:PV}) and used 
in \cite{Blaschke:2008wf,Blaschke:2005hs} with the amplitudes $f_{(0)}$ and 
$f_{(2)}$ as the results of the numerical analysis.

The $E_{0}^{2}$ dependence of $f_{\rm max}$  and $n_{\rm max}$ 
(see Fig.~\ref{fig:nmax-E0}) follows immediately from Eqs.~(\ref{12}) and 
(\ref{13}).
The time dependent component (\ref{13}) does not depend on $\nu$ for
$\nu \ll m$ (see Fig.~\ref{fig:landscape}, calm valley); 
the constant component (\ref{12}) brings some small dependence in this domain.
The calculations close to the poles in Eqs.~(\ref{12}) and (\ref{13}) require a
more careful analysis. 
The Eqs.~(\ref{12}) and (\ref{13}) indicate the one-photon mechanism 
of the EPP bubble generation in the resonance domain $\nu \simeq 2\omega_0 
(\ve{p})$ (see Sect.~\ref{sec:PV}, Subsect.~\ref{ssec:res}). 
In the considered approximation this bubble has a spherical form.

The estimates of the density (\ref{9}) on the basis of Eqs.~(\ref{12}) and
(\ref{13}) lead to the result
\begin{equation}
n(t) = n_{(0)} + \frac{1}{2} n_{(2)}[1 - \cos (2\nu t)]~.
\label{14}
\end{equation}
In the leading approximation with respect to small $\nu/\omega_0$ it follows 
from Eqs.~(\ref{12}) and (\ref{13}) that
\begin{eqnarray}
\label{15}
n_{(0)} &=& \frac{5}{3} \frac{\alpha}{m} \left(\frac{E_{0}}{16}\right)^2
\biggl(\frac{\nu}{m}\biggr)^2~,\\
n_{(2)}&=&\frac{3}{2} \frac{\alpha}{m} \left(\frac{E_{0}}{4}\right)^2~.
\label{15a}
\end{eqnarray}
The ratio of these expressions is equal to 
$n_{(0)}/n_{(2)} = (5/72) (\nu/m)^2$ , 
which is close to the estimate given in \cite{Blaschke:2005hs}.
The frequency independence and $E_0^2$ dependence of the amplitude
$n_{(2)}$ of the number density oscillations agrees with  the
numerical calculations exhibited in Fig.~\ref{fig:landscape}.
The constant (residual) component density (\ref{15}) was not observed 
previously in numerical calculations.
The numerical integration shows that the estimates (\ref{15}) and (\ref{15a}) 
remain valid up to the quasiclassical boundary $\nu \lesssim 2m$.
The asymptotics  at $\nu \to 0$ in Eqs.~(\ref{12}) and (\ref{15}) leads to a 
vanishingly small residual EPP density.

Thus, in the multiphoton domain of the calm valley the averaged production 
rate $<\dot{f}>_T$ of EPP is negligibly small. 
Instead, there are vacuum oscillations induced by the external field and the 
small constant (residual) component. 
In principle, these vacuum effects can lead to some observable effects. 
For instance, they influence on the propagation of an electromagnetic probe
signal. This can result in an anomalous absorption of this signal at the 
frequency $2 \nu$.

The perturbation theory used here in the quasiparticle representation is valid 
in the multiphoton domain $\gamma \gg 1$, when the swing amplitude 
(Sect.~\ref{sec:PV}) is small. 
However, the numerical solutions of the KE (\ref{ke}) show that these results 
are valid also in some parts of the tunneling mechanism region. 
At first glance, this fact is unexpected. 
Indeed, in this region for subcritical fields $E_0 \ll E_c$ it is very well 
known that the EPP production rate is exponentially small 
\cite{Grib:1994,Brezin:1970xf}. 
But this effect is not accessible to perturbation theory. 
On the other hand, the domain of the long wavelength limit 
$\lambda \gtrsim \lambda_0$ is difficult for the numerical analysis and it was 
not studied in the present work. 
It is possible, that some special solution of KE (\ref{ke}) is present here 
which is compatible with the well known estimates 
\cite{Grib:1994,Brezin:1970xf}.

\section{Summary} 
\label{sec:sum}

In the present work we have described the PV response to a monochromatic laser 
radiation in the landscape of the laser parameters (\ref{01}) and (\ref{02}). 
Three specific domains were observed here
\begin{itemize}
  \item[-] the domain of the vacuum oscillations induced by the external field 
(calm valley), where the EPP production rate is very small, 
$<\dot{f}>_T\approx 0$, and the breathing modes are acting only, 
  \item[-] the domain of the accumulation effect due to strong field (F), and
  \item[-] the resonant (R) accumulation domain, where the EPP production rate 
can reach significant values.
\end{itemize}

Thus, it was shown that the behavior of the PV in the calm valley up to a 
certain critical point $\lambda_{acc}= \pi \lcbar$ is stable and does not 
depend on the wavelength $\lambda$ of the external electromagnetic field.
This domain accommodates different mechanisms of vacuum decay into 
quasiparticle EPP excitations
\begin{itemize}
\item[-] tunneling ($\gamma\ll 1$, when the photon number of the external
field is infinity or very large), 
\item[-] multiphoton processes ($\gamma \ge 1$) and 
\item[-] few-photon processes.
 \end{itemize}

The accumulation effect depicted in Fig.~\ref{fig:landscape}, appears just 
either in the multiphoton domain as a parametric resonance induced by the 
one-photon mechanism of EPP creation at the frequency $\nu = 2m$  (domain R) 
or as a result of the acceleration of quasiparticle $e^-e^+$ pairs in a 
subcritical external field (domain F).
These processes are accompanied by the accumulation of EPP density and are
limited by the condition $f(\mathbf{p},t) \le 1$. 
For increasing $E_0$, the domain of the momentum space covered by the 
distribution  function is enlarged too  ("broadening" of $f(\mathbf{p},t)$,
see Fig.~\ref{degen}). 
This can lead to a rapid growth of the EPP density.

For the strong and moderately strong fields (see Fig.~\ref{fig:landscape}) 
these predictions are based on a realistic KE (\ref{ke}) in the scope of its 
applicability. 
For sufficiently small fields, $E_0 \ll E_c$, the accumulation R-domain comes 
in the region of the strongly fluctuating external fields, where the KE 
(\ref{ke}) is not applicable, generally speaking, and its solutions have here 
only exploratory character.

As the EPP density under the accumulation conditions in the R domain can be
very high  even  for relatively weak fields, $E_0 \ll E_c$, it can be
expected that the manifestation of the dynamical Schwinger effect can be quite
possible in  this  domain  due  to the generation of secondary
effects such as, e.g., the radiation of annihilation photons.
Apparently, these remarks can be useful for the discussion of possible
experiments for observing the dynamical Schwinger effect.
On the other hand,  the accumulation effect in the both accumulation domains 
R and F in the short-wavelength domain can be considerable only for a 
sufficiently long duration of the laser field action $t \gg T$ 
(Figs.~\ref{fig_accum}, \ref{degen} and \ref{gauss}). 
This circumstance makes difficult an experimental observation of the effect 
in the case of a short single laser pulse.

In the theory of vacuum EPP creation under the action of a periodic laser field
of the type (\ref{field}) the residual EPP density is the basic final product 
of the theory. 
However, its definition can be difficult. 
In the present work the residual density is understood, as a rule, as the EPP 
density at the final moment of action of some field periods 
(see, e.g., \cite{Grib:1994,Brezin:1970xf}). 
It corresponds to the density $n_{(0)}$ (\ref{15}). 
In such a case it is implied that the field (\ref{field}) either prolongs to 
act farther than the periodical field or it switches off instantly. 
Both these cases are rather artificial. 
It would be more satisfactory to consider some pulsed field of the type 
(\ref{10a}), when a periodical field switches on and off gradually or 
adiabatically. 
The similar situation was discussed in the Sect.~\ref{sec:PV}. 
It is very important that it does not lead to qualitative change in the 
picture for moderate fields thus demonstrating the stability of the obtained 
results with respect to the way of switching the laser field on and off.

Let us underline also that the approach used in the present work is based on 
a minimal number of very general assumptions (spatial homogeneity of the 
external field and its linear polarisation) and is in essence an exact 
consequence of the basic QED in the framework of these limitations. 
This circumstance allows the authors to express the opinion that the obtained 
results in the framework of the given approach are free from some additional 
approximation (e.g., of the WKB type) and therefore a more trustworthy in the
QC-domain. Moreover, they allow to make first prognoses for the Q-domain.
On the other hand, the basic kinetic equation (KE) (\ref{ke}) is valid in the
case of a quasiclassical electric field (\ref{field}) only.
We extrapolated the domain of its applicability into the Q domain.
The foundation for such an extrapolation is the assumption that the solution
of the KE (\ref{ke}) in the boundary domain is valid approximately because
there the quantum fluctuations of the electric field are not too large.
Thus, we observe here the tendency to a non-monotonicity described above.
In the Q domain the external field can be described  with the help  of
the  corresponding density matrix of the electromagnetic  field.
Then, the quasiclassical field $A^{\mu}$ will play the role of a
background field, i.e. $A_{tot}^{\mu} = A^{\mu} + \hat{A}^{\mu}$,  where
$\hat{A}^{\mu}$  is  the  field operator of the photon component of the total
field  $A_{tot}^{\mu}$.

For a more adequate investigation of the PV response in the  Q domain it is
necessary to develop a kinetic theory for the generalization of the KE
(\ref{ke}) taking into account the quantum fluctuations of the electric field.
For understanding the situation, we mention the works
\cite{Krive:1984xi,Popov:2002} in which the
vacuum particle creation is considered under the influence of a stochastic
time dependent electric field.
We want to mention two of the possibilities which emerge here.
The compressed state of the electromagnetic field (e.g., \cite{Bykov:1991})
is a quantum one but contains a large occupation number of photons.
As a result, the criterion (6) of the quasiclassical case can be not adequate
to this situation.
For the second example we remark that, in the case of a rather
strong field $E_{0} \gtrsim 0.1~E_{c}$,  the backreaction mechanism
becomes essential and leads to a stochastic (or close to one)
internal electric field \cite{Vinnik:2001qd}.
Thus, the kinetic description of vacuum particle creation in the Q domain is
an actual problem deserving further investigation.

Considering the muon PV independently of $e^-e^+$ PV, it can be
expected that the response of the $\mu^-\mu^+$ PV will repeat on the
qualitative level the picture described in Sect.~\ref{sec:PV} with a shift 
to the side of shorter waves.

Two regions in the plane of the parameters (\ref{01}) and (\ref{02}) 
remain uninvestigated in the framework of the used approach: 
the top left corner on Fig.~\ref{fig:landscape} where short 
wavelengths and strong fields simultaneously persist, and the right boundary 
of the tunneling domain in the calm valley where $\lambda > \lambda_0$. 
It would be worthwhile to investigte these areas separately.
Preliminary results of this work have been reported recently
\cite{Dresden2011} and a more elaborate discussion is in preparation. 

\section{Conclusion}

We have investigated the behavior of the quasiparticle EPP generated
from the PV under the action of a strong laser field.
It was shown that particle production in the initial stage of the field 
action is characterized by an equilibrium like ``thermal''  distribution.  
However, at later times the quasiparticle EPP distribution becomes very 
complicated and shows a far-from-equilibrium  behaviour with distinct features
depending on the specific domain of the landscape.
     In  the  subsequent,  second  part  of the work we plan to investigate
features  of  the  real EPP which remains after the laser pulse ceases. 
Moreover, the transient phenomenona between quasiparticle and residual
states of the EPP shall be studied.

\subsection*{Acknowledgements}
We are grateful to A.~M.~Fedotov, L.~Juchnowski, A.~G.~Lavkin, 
A.~V.~Tarakanov and V.~D.~Toneev for useful discussions on different aspects 
of this work.
The work of D.B.B. was supported in part by the Russian Fund for Basic
Research under grant No. 11-02-01538-a.
S.A.S. and S.M.S. acknowledge support by Deutsche Forschungsgemeinschaft 
(DFG) under Project number TO 169/16-1.



\begin{thebibliography}{99}

\bibitem{Baur:2007df}
  G.~Baur,
  Nucl.\ Phys.\ Proc.\ Suppl.\  {\bf 184}, 143 (2008).

\bibitem{Dodonov:2010zza}
  V.~V.~Dodonov,
  Phys.\ Scripta {\bf 82}, 038105 (2010).

\bibitem{Kluger:1992}
Y.~Kluger, J.~M.~Eisenberg, B.~Svetitsky, F.~Cooper and E.~Mottola,
Phys. Rev. D {\bf 45}, 4659 (1992).

\bibitem{Schmidt:1998vi}
S.~M.~Schmidt, D.~Blaschke, G.~R\"opke, S.~A.~Smolyansky, A.~V.~Prozorkevich
and V.~D.~Toneev,
Int.\ J.\ Mod.\ Phys.\ E \textbf{7}, 709 (1998).

\bibitem{Pervushin:2006vh}
  V.~N.~Pervushin and V.~V.~Skokov,
  Acta Phys.\ Polon.\  {\bf B37}, 2587 (2006).

\bibitem{Blaschke:2005hs}
  D.~B.~Blaschke, A.~V.~Prozorkevich, C.~D.~Roberts, S.~M.~Schmidt and
  S.~A.~Smolyansky,
  Phys.\ Rev.\ Lett.\  {\bf 96}, 140402 (2006).

\bibitem{Blaschke:2009uy}
  D.~B.~Blaschke, S.~M.~Schmidt, S.~A.~Smolyansky and A.~V.~Tarakanov,
  Phys.\ Part.\ Nucl.\  {\bf 41}, 1004 (2010).

\bibitem{Blaschke:2010vs}
  D.~B.~Blaschke, G.~R\"opke, S.~M.~Schmidt, S.~A.~Smolyansky and 
  A.~V.~Tarakanov,
  Contrib.\ Plasma Phys.\  {\bf 51}, 451 (2011).

\bibitem{Smolyansky:2010as}
S.~A.~Smolyansky, D.~B.~Blaschke, A.~V.~Chertilin, G.~R\"opke and 
A.~V.~Tarakanov,
  [arXiv:1012.0559 [physics.plasm-ph]].

\bibitem{Blaschke:2011af}
 D.~B.~Blaschke, G.~R\"opke, V.~V.~Dmitriev, S.~A.~Smolyansky and 
A.~V.~Tarakanov,
  arXiv:1101.6021 [physics.plasm-ph].
			
\bibitem{Blaschke:2011is}
  D.~B.~Blaschke, V.~V.~Dmitriev, G.~R\"opke and  S.~A.~Smolyansky,
  Phys.\ Rev.\ D {\bf 84}, 085028 (2011).

\bibitem{DiPiazza:2006pr}
  A.~Di Piazza, K.~Z.~Hatsagortsyan and C.~H.~Keitel,
  Phys.\ Rev.\ Lett.\  {\bf 97}, 083603 (2006).

\bibitem{Sauter:1931zz}
  F.~Sauter,
  Z.\ Phys.\  {\bf 69}, 742 (1931).

\bibitem{Heisenberg:1935qt}
  W.~Heisenberg and H.~Euler,
  Z.\ Phys.\  {\bf 98}, 714 (1936).

\bibitem{Schwinger:1951nm}
  J.~S.~Schwinger,
  Phys.\ Rev.\  {\bf 82}, 664 (1951).

\bibitem{Greiner:1985ce}
  W.~Greiner, B.~M\"uller and J.~Rafelski,
  "Quantum Electrodynamics Of Strong Fields,"
  Berlin, Germany: Springer (1985).

\bibitem{Grib:1994}
A.~A.~Grib, S.~G.~Mamaev and V.~M.~Mostepanenko,
{\it Vacuum Quantum Effects in Strong External Fields},
Friedman Laboratory Publish., St.~Petersburg (1994).

\bibitem{Brezin:1970xf}
  E.~Brezin and C.~Itzykson,
  Phys.\ Rev.\  D {\bf 2}, 1191 (1970).

\bibitem{Bulanov:2004de}
N.~B.~Narozhny, S.~S.~Bulanov, V.~D.~Mur and V.~S.~Popov,
  Phys.\ Lett.\  A {\bf 330}, 1 (2004); 
  JETP Lett. {\bf 80}, 382 (2004) 
  [Pis'ma Zh. Eksp. Teor. Fiz. {\bf 80}, 434 (2004)].

\bibitem{Narozhny:2006}
N.~B.~Narozhny, S.~S.~Bulanov, V.~D.~Mur and V.~S.~Popov,
Zh. Eksp. Teor. Fiz. {\bf 129}, 14 (2006) [JETP {\bf 102}, 9 (2006)].

\bibitem{Fedotov:2006}
A.~M.~Fedotov, Laser Phys. {\bf 19}, 214 (2009).

\bibitem{Blaschke:2008wf}
  D.~B.~Blaschke, A.~V.~Prozorkevich, G.~R\"opke, C.~D.~Roberts, S.~M.~Schmidt,
D.~S.~Shkirmanov and S.~A.~Smolyansky,
  Eur.\ Phys.\ J.\ D {\bf 55}, 341 (2009).

\bibitem{Filatov:2006}
A.~V.~Filatov, A.~V.~Prozorkevich and S.~A.~Smolyansky,
Proc. of SPIE {\bf 6165}, 616509 (2006).

\bibitem{Filatov:2009xd}
  A.~V.~Filatov, S.~A.~Smolyansky and A.~V.~Tarakanov,
 Proc. of the XX International Baldin Seminar on High Energy Physics Problems 
 "Relativistic Nuclear Physics and Quantum Chromodynamics",
 Dubna, Sept. 29 - Oct. 04, 2008, 202 (2008).

\bibitem{Berestetskii}
V.~B.~Berestetskii, E.~M.~Lifshits and L.~P.~Pitaevskii,
{\it Quantum Electrodynamics}, Pergamon, New York (1982).

\bibitem{Ritus:1979} V.I. Ritus, 
Tr. Fiz. Inst. Akad. Nauk SSSR \textbf{111}, 5 (1979) (in Russian).

\bibitem{Popov:2001}
V.~S.~Popov, Sov. Phys. JETP {\bf 120}, 315 (2001).

\bibitem{Popov:2004}
V.~ S. Popov,
Phys. Usp.  \textbf{47}, 855  (2004).

\bibitem{Delone:1998}
N.~B.~Delone, V.~P.~Krainov,
Phys. Usp.  {\bf 41}, 469 (1998).

\bibitem{Blaschke:2010sgu}
D.~B.~Blaschke, V.~V.~Dmitriev, P.~I.~Smolyansky, S.~A.~Smolyansky and
A.~V.~Chertilin, 
Proceedings of Saratov University, Ser. Physics, {\bf 10}, 45 (2010) 
(in Russian).

\bibitem{Liesfeld:2005}
B.~Liesfeld, J.~Bernhardt, K.~U.~Amthor, H.~Schwoerer and R.~Sauerbrey,
Appl. Phys. Lett. {\bf 86}, 161107 (2005).

\bibitem{Gregori:2010uf}
 G.~Gregori, D.~B.~Blaschke, P.~P.~Rajeev, H.~Chen, R.~J.~Clarke, T.~Huffman,
C.~D.~Murphy, A.~V.~Prozorkevich, C.~D.~Roberts, G.~R\"opke, S.~M.~Schmidt,
S.~A.~Smolyansky, S.~Wilks and R.~Bingham,
  High Energy Dens.\ Phys.\  {\bf 6}, 166 (2010).

\bibitem{Ringwald:2001ib}
  A.~Ringwald,
  Phys.\ Lett.\  {\bf B510}, 107 (2001).

\bibitem{DiPiazza:2011tq}
  A.~Di Piazza, C.~M\"uller, K.~Z.~Hatsagortsyan and C.~H.~Keitel,
  Rev.\ Mod.\ Phys.\  {\bf 84}, 1177 (2012).

\bibitem{Korzhimanov:2011}
A.~V. Korzhimanov, A.~A. Gonoskov, E.~ A. Khazanov and A.~M. Sergeev,
Phys. Usp. {\bf 54}, 9 (2011).

\bibitem{Ruffini:2009hg}
  R.~Ruffini, G.~Vereshchagin and S.~-S.~Xue,
  Phys.\ Rept.\  {\bf 487}, 1 (2010).

\bibitem{Schmidt:1998zh}
  S.~M.~Schmidt, D.~Blaschke, G.~R\"opke, A.~V.~Prozorkevich, S.~A.~Smolyansky 
 and V.~D.~Toneev,
  Phys.\ Rev.\ D {\bf 59}, 094005 (1999).

\bibitem{854971}
  G.~Breit and J.~A.~Wheeler,
  Phys.\ Rev.\ \ {\bf 46}, 1087  (1934).

\bibitem{Hebenstreit:2011wk}
  F.~Hebenstreit, R.~Alkofer and H.~Gies,
  Phys.\ Rev.\ Lett.\  {\bf 107}, 180403 (2011).

\bibitem{Gies:2005bz} 
  H.~Gies and K.~Klingm\"uller,
  Phys.\ Rev.\ D {\bf 72}, 065001 (2005).

\bibitem{Dunne:2006ur} 
  G.~V.~Dunne and Q.~-H.~Wang,
  Phys.\ Rev.\ D {\bf 74}, 065015 (2006).

\bibitem{Nikishov:1969tt}
  A.~I.~Nikishov,
  Zh.\ Eksp.\ Teor.\ Fiz.\  {\bf 57}, 1210 (1969) (in Russian).

\bibitem{Narozhnyi:1970uv}
  N.~B.~Narozhnyi and A.~I.~Nikishov,
  Yad.\ Fiz.\  {\bf 11}, 1072 (1970)
  [Sov.\ J.\ Nucl.\ Phys.\  {\bf 11}, 596 (1970)].

\bibitem{Filatov:2007ha}
  A.~V.~Filatov, A.~V.~Prozorkevich, S.~A.~Smolyansky and V.~D.~Toneev,
  Phys.\ Part.\ Nucl.\  {\bf 39}, 886 (2008).

\bibitem{Castorina:2007eb}
  P.~Castorina, D.~Kharzeev and H.~Satz,
  Eur.\ Phys.\ J.\ C {\bf 52}, 187 (2007).

\bibitem{Alkofer:2001ik} 
  R.~Alkofer, M.~B.~Hecht, C.~D.~Roberts, S.~M.~Schmidt and D.~V.~Vinnik,
  Phys.\ Rev.\ Lett.\  {\bf 87}, 193902 (2001).

\bibitem{Roberts:2002py}
  C.~D.~Roberts, S.~M.~Schmidt and D.~V.~Vinnik,
  Phys.\ Rev.\ Lett.\  {\bf 89}, 153901 (2002).

\bibitem{Schutzhold:2008pz} 
  R.~Sch\"utzhold, H.~Gies and G.~Dunne,
  Phys.\ Rev.\ Lett.\  {\bf 101}, 130404 (2008).

\bibitem{Orthaber:2011cm} 
  M.~Orthaber, F.~Hebenstreit and R.~Alkofer,
  Phys.\ Lett.\ B {\bf 698}, 80 (2011).

\bibitem{Fey:2011if} 
  C.~Fey and R.~Sch\"utzhold,
  Phys.\ Rev.\ D {\bf 85}, 025004 (2012).

\bibitem{Krive:1984xi}
  I.~V.~Krive and L.~A.~Pastur,
  Yad.\ Fiz.\  {\bf 39}, 224 (1984) (in Russian).

\bibitem{Popov:2002}
A.~M.~Popov and O.~V.~Tikhonova,  
JETP \textbf{95}, 844 (2002).

\bibitem{Bykov:1991}
 V.~P.~Bykov,
  Sov. Phys. Usp. {\bf 34},  910 (1991).

\bibitem{Vinnik:2001qd}
  D.~V.~Vinnik, A.~V.~Prozorkevich, S.~A.~Smolyansky, V.~D.~Toneev, 
  M.~B.~Hecht, C.~D.~Roberts and S.~M.~Schmidt,
  Eur.\ Phys.\ J.\ C {\bf 22}, 341 (2001).

\bibitem{Dresden2011}
S.~A.~Smolyansky (in collaboration with D.~B.~Blaschke, A.~M.~Fedotov,
A.~G.~Lavkin and A.~V.~Prozorkevich),
Workshop "Petawatt Lasers and Hard X-Ray Light Sources", Dresden-Rossendorf,
Sept. 5 to 9,  2011; http://www.hzdr.de/PWLasers2011.

%

	
\end{thebibliography}
\end{document}